\documentclass[prd,twocolumn,showpacs,floatfix,amsmath,nofootinbib,amssymb,floatfix]{revtex4}
\usepackage{graphicx,color,dcolumn,booktabs,bm,multirow}
\usepackage{longtable,lscape}
\usepackage{txfonts}
\usepackage{enumitem}
\usepackage{overpic}
\usepackage{amssymb}
\usepackage{indentfirst}
\usepackage{feynmf}   
\usepackage{slashed}  
\usepackage{cases}
\usepackage{color}
\usepackage{multirow}
\usepackage{epstopdf}
\usepackage{graphicx,color,dcolumn,booktabs,bm}
\usepackage[colorlinks,
citecolor=blue,
anchorcolor=red,
menucolor=red,
linkcolor=red,
filecolor=red,
runcolor=red,
urlcolor=blue,
frenchlinks=red]{hyperref}
\usepackage{amsmath}
\usepackage{mathrsfs}
\usepackage{epsfig}
\usepackage{youngtab}
\usepackage{graphicx}
\usepackage{txfonts}
\usepackage{cleveref}

\begin{document}
\title{Probing the nature of the anticharmed-strange pentaquark states: mass spectra, decays, and magnetic moments}

\author{Xuejie Liu$^1$}\email[E-mail: ]{1830592517@qq.com}
\author{Yue Tan$^{2}$}\email[E-mail:]{tanyue@ycit.edu.cn}
\author{Xiaoyun Chen$^{5}$}\email[E-mail:]{xychen@jit.edu.cn}
\author{Dianyong Chen$^{3,6}$\footnote{Corresponding author}}\email[E-mail:]{chendy@seu.edu.cn}
\author{Hongxia Huang$^4$}\email[E-mail:]{hxhuang@njnu.edu.cn}
\author{Jialun Ping$^4$}\email[E-mail: ]{jlping@njnu.edu.cn}
\affiliation{$^1$School of Physics, Henan Normal University, Xinxiang 453007, P. R. China}
\affiliation{$^2$School of Mathematics and Physics, Yancheng Institute of Technology, Yancheng, 224051,  P. R. China}
\affiliation{$^3$Lanzhou Center for Theoretical Physics, Lanzhou University, Lanzhou 730000, P. R. China}
\affiliation{$^4$Department of Physics, Nanjing Normal University, Nanjing 210023, P. R. China}
\affiliation{$^5$College of Science, Jinling Institute of Technology, Nanjing 211169, P. R. China}
\affiliation{$^6$School of Physics, Southeast University, Nanjing 210094, P. R. China}

\begin{abstract}
Within the framework of the quark delocalization color screening model, a systematic investigation of the anticharmed-strange pentaquark system is performed using the resonance group method. The currently estimations predict three bound states with estimated masses to be 2886 MeV, 3039 MeV, and 3153 MeV, respectively. Additionally, three resonance states are identified in various scattering phase shifts processes. Among them, two resonance states $\Sigma D$ and $\Sigma^{\ast}D^{\ast}$ with quantum number $\frac{1}{2}(\frac{1}{2}^{-})$ are detected in channels $ND_{s}^{\ast}$ and $ND$, and $\Sigma D^{\ast}$ and $\Lambda D$, with masses and decay widths of ($M_{R}=3053\sim3055$ MeV, $T_{total}=13.0\sim13.4$ MeV) and ($M_{R}=3389\sim3390$ MeV, $T_{total}=10.4$ MeV), respectively. In the $\Lambda D^{\ast}$ and $\Sigma D^{\ast}$ channels, a resonance state with quantum number $\frac{1}{2}(\frac{3}{2}^{-})$ is discovered, with its mass and decay width being $3250\sim3252$ MeV and 4.4 MeV, respectively. These predicted pentaquark states have $\bar{c}snnn$ quark compositions,  allowing them to be recognized as genuine pentaquark states. To validate these predictions, it is expected that upcoming experiments will further explore the predicted resonance and bound states in these possible decay channels.

\end{abstract}

\maketitle

\section{\label{sec:introduction}Introduction}
Since the proposal of the quark model~\cite{Gell-Mann:1964ewy,Jaffe:1976yi}, the concept of exotic hadron states has also emerged, garnering significant attention from both theorists and experimentalists. With the continuous accumulation of experimental data,  more exotic hadron states have been observed, including charmonium-like states~\cite{Belle:2003nnu,BaBar:2005hhc,BESIII:2013ris,BESIII:2013ouc,Belle:2011aa}, hidden-charm pentaquarks~\cite{LHCb:2015yax,LHCb:2019kea,LHCb:2020jpq,LHCb:2022ogu}, doubly charmed tetraquarks~\cite{LHCb:2021vvq,LHCb:2021auc}, and charmed strange tetraquarks~\cite{LHCb:2020bls,LHCb:2020pxc,LHCb:2022sfr,LHCb:2022lzp}, which enriches the hadron family. However, due to limited experimental data and the reliance on specific models, the nature of these states still lacks a definitive consensus, so the identification and investigations of these states' properties have remained core subjects in hadron physics. Researching exotic hadron states not only helps us understand the non-perturbative behavior of strong interactions in QCD but also reveals the internal structural patterns of exotic states. Thus, up to now, the quest for exotic states remains an important ongoing research area in hadron physics.

In recent years, with the advancement of experimental techniques, especially near hadron thresholds, an increasing number of exotic hadronic states have been discovered. The exploration of hidden-charm pentaquarks states, particularly since the LHCb  Collaboration announced the discovery of $P_{c}(4312)$,  $P_{c}(4380)$,  $P_{c}(4440)$,  $P_{c}(4457)$,  $P_{c}(4459)$,  $P_{c}(4337)$,  $P_{cs}(4459)$ and  $P_{cs}(4338)$, have gained the significant breakthroughs for hadron physics~\cite{LHCb:2015yax,LHCb:2019kea,LHCb:2020jpq,LHCb:2022ogu}. These findings have inspired theorists to employ various methods and models, proposing multiple interpretations, including compact pentaquark states~\cite{Ali:2019clg,Wang:2019got,Ali:2019npk}, loosely bound molecular states~\cite{Chen:2019bip,Liu:2019tjn,Shimizu:2019ptd,Wu:2019rog,Lin:2019qiv,Peng:2020hql,Xiao:2021rgp,Lu:2021irg,Zhu:2022wpi,Azizi:2023iym}, and kinematical effects~\cite{Guo:2015umn,Nakamura:2021qvy,Liu:2015fea}. The molecular picture is gaining increasing recognition in the communities, as the aforementioned exotic states are very close to the threshold of a pair of conventional hadrons. Nonetheless, there are still many controversial arguments about the properties of the hidden-charm pentaquark states. From above, identifying genuine exotic hadronic states is challenging, therefore, it is crucial to search for experimentally apparent exotic hadronic states to construe a new family of hadron physics.

Very recently, the LHCb Collaboration utilized proton-proton collision data for the first time to study the $\Lambda_{b}^{0}\rightarrow D^{+}D^{-}\Lambda$ decay at nearly 13 TeV energy levels~\cite{LHCb:2024hfo}. By inspecting the invariant mass distributions of $\Lambda D^{+}$, abundant intermediate resonances may be discovered in the decay process. These hints imply the possible existence of open-charm pentaquark states with quark content $\bar{c}snnn$ (n: u or d). If this state is validated, it will undoubtedly signify a unique pentaquark state.

Indeed, we have noticed that several early studies have already explored the $\bar{c}snnn$ pentaquark states. In Refs.~\cite{Lipkin:1987sk,Gignoux:1987cn,Karl:1987uf}, a doublet of states, the $P^{0}_{\bar{c}s}$ and $P^{-}_{\bar{c}s}$, and their charge conjugate states with spin 1/2  were considered more likely to be stable pentaquark states than  a $S=0$ charmed pentaquark ($\bar{c}qqqq$) due to the lack of a quark exchange process in the lowest decay channel $ND_{s}^{-}$ in the constituent quark model based on one-gluon exchange interaction between quarks.  A systematic study of all pentaquarks was performed in Ref.~\cite{Leandri:1989su} where over a dozen are found to be candidates for stability. Among them, those with strangeness $S=-1$ or $-2$ were the most favorable.  In the diquark picture, Ref.~\cite{Stewart:2004pd} estimated the $T_{s}(\bar{c}snnn)$ mass to be about 2580 MeV. However, the theoretical mass of $D_{s}p$ is 2910 MeV during $T_{s}\rightarrow D_{s}p$, which is about 330 MeV above the $T_{s}$ mass. This finding reveals a new pentaquark state and indicates relative stability against strong decay. In addition, the instanton model~\cite{Takeuchi:1993rs}, Skyrme model~\cite{Riska:1992qd,Oh:1994np,Oh:1994yv,Chow:1994hg,Chow:1995hv}, and the Chromomagnetic Interaction model~\cite{An:2020vku} also predicted bound open-charm pentaquark states with $S=-1$. Nevertheless, some theoretical studies have yielded results to the contrary. For example, in Refs.~\cite{Fleck:1989ff,Zouzou:1993ri,Genovese:1997tm,Stancu:1998sm}, the calculations indicated that $S=-1$ open-charm pentaquarks were probably unbound. Experimentally,  based on the consequences of Refs.~\cite{Lipkin:1987sk,Gignoux:1987cn}, a search is made for the decay $P^{0}_{\bar{c}s}\rightarrow \phi\pi p$ and $P^{0}_{\bar{c}s}\rightarrow K^{\ast 0}K^{-}p$ in data from Fermilab E791 Collaboration, but no evidence is found~\cite{E791:1997xzq,E791:1999pcs}. Though the results from different approaches with various inner structures are disaccord with each other, the difference may directly verify the innner structure once the states are discovered in the future.

The aim of the present work is to investigate the masses, decay properties, and magnetic moments of open-charm pentaquarks with strangeness. First, we systematically evaluate the effective potential at different quantum numbers within the framework of a quark delocalization color screening model (QDCSM), followed by dynamical estimation of the open-charm pentaquarks with strangeness through the resonance group method and channel coupling effect to search for possible bound states. Meanwhile, with the consideration of quantum number conservation and phase space constraints, we probe the hadronic decay process of the open-charm pentaquarks with strangeness to identify possible resonance states. In addition, the magnetic moment of the pentaquark states, which is a measure of its ability to interact with the magnetic field, is crucial for understanding the behavior of these particles. Therefore, we directly computed the magnetic moments of the possible pentaquark states using flavor-spin wave functions to acquire information about their magnetic moments.

This work is organized as follows. In the next section, the detail of the QDCSM is presented.  In section~\ref{dis}, a comprehensive numerical analysis of the open-charm pentaquark with strangeness is carried out, including the effective potential, possible bound states, resonance states, and magnetic moments. Finally, section~\ref{sum} summarizes the main conclusions of this present work.

\section{THE QUARK DELOCALIZATION COLOR SCREENING MODEL  \label{mod}}

The QDCSM is an extension of the native quark cluster model~\cite{DeRujula:1975qlm,Isgur:1979be,Isgur:1978wd,Isgur:1978xj} and was developed with the aim of
addressing multiquark systems (More detail of QDCSM can be found in the Refs.~\cite{Wang:1992wi,Chen:2007qn,Chen:2011zzb,Wu:1996fm,Huang:2011kf}).
In the QDCSM, the general form of the Hamiltonian for the pentaquark system is,
\begin{equation}
H = \sum_{i=1}^{5} \left(m_i+\frac{\boldsymbol{p}_i^2}{2m_i}\right)-T_{\mathrm{CM}}+\sum_{j>i=1}^5V(r_{ij}),\\
\end{equation}
where the center-of-mass kinetic energy, $T_{\mathrm{CM}}$, is subtracted without losing generality since we mainly focus on the internal relative motions of the multiquark system. The two body potentials include the color-confining potential, $V_{\mathrm{CON}}$, one-gluon exchange potential, $V_{\mathrm{OGE}}$, and Goldstone-boson exchange potential, $V_{\chi}$, respectively, i.e.,
\begin{equation}
V(r_{ij}) = V_{\mathrm{CON}}(r_{ij})+V_{\mathrm{OGE}}(r_{ij})+V_{\chi}(r_{ij}).
\end{equation}

Noted herein that the potentials include the central, spin-spin, spin-orbit, and tensor contributions, respectively. Since the current calculation is based on S-wave, only the first two kinds of potentials will be considered attending the goal of the present calculation and for clarity in our discussion. In particular, the one-gluon-exchange potential, $V_{\mathrm{OGE}}(r_{ij})$, reads,
\begin{eqnarray}
V_{\mathrm{OGE}}(r_{ij}) &=& \frac{1}{4}\alpha_{s} \boldsymbol{\lambda}^{c}_i \cdot\boldsymbol{\lambda}^{c}_j \nonumber\\
&&\times\left[\frac{1}{r_{ij}}-\frac{\pi}{2}\delta(\boldsymbol{r}_{ij})\left(\frac{1}{m^2_i}+\frac{1}{m^2_j}
+\frac{4\boldsymbol{\sigma}_i\cdot\boldsymbol{\sigma}_j}{3m_im_j}\right)\right],\ \
\end{eqnarray}
where $m_{i}$ is the quark mass, $\boldsymbol{\sigma}$ and $\boldsymbol{\lambda^{c}}$ are the Pauli matrices and SU(3) color matrices, respectively. The QCD-inspired effective scale-dependent strong coupling constant, $\alpha_{s}$, offers a consistent description of mesons and baryons from the light to the heavy quark sectors, which can be written by,
\begin{equation}
\alpha_{s}(\mu)= \frac{\alpha_{0}}{\ln(\frac{\mu^{2}+\mu_{0}^{2}}{\Lambda_{0}^2})}.
\end{equation}

In the QDCSM, the confining interaction $V_{\mathrm{CON}}(r_{ij})$ can be expressed as
\begin{equation}
 V_{\mathrm{CON}}(r_{ij}) =  -a_{c}\boldsymbol{\lambda^{c}_{i}\cdot\lambda^{c}_{j}}\Big[f(r_{ij})+V_{0_{ij}}\Big] \ ,
\end{equation}
where $a_{c}$ represents the strength of the confinement potential and $V_{0_{ij}}$ refers to the zero-point potential. Moreover, in the quark delocalization color screening model, the quarks in the considered pentaquark state $\bar{c}snnn$ are first divide into two clusters, which are baryon cluster composed of three quarks, and meson cluster composed of one quark and one antiquark. And then the five-body problem can be simplified as a two-body problem the $f(r_{ij})$ is,
\begin{equation}
 f(r_{ij}) =  \left\{ \begin{array}{ll}r_{ij}^2 & \quad \mbox{if }i,j\mbox{ occur in the same cluster}, \\
\frac{1 - e^{-\mu_{ij} r_{ij}^2} }{\mu_{ij}} & \quad \mbox{if }i,j\mbox{ occur in different cluster},
\end{array} \right.
\label{Eq:fr}
\end{equation}
where the color screening parameter $\mu_{ij}$ is determined by fitting the deuteron properties, nucleon-nucleon, and nucleon-hyperon scattering phase shifts~\cite{Chen:2011zzb,Wang:1998nk}, with $\mu_{nn}= 0.45\ \mathrm{fm}^{-2}$ , $\mu_{ns}= 0.19\ \mathrm{fm}^{-2}$
and $\mu_{ss}= 0.08\ \mathrm{fm}^{-2}$, satisfying the relation $\mu_{ns}^{2}=\mu_{nn}\mu_{ss}$, where $n$ represents $u$ or $d$ quark. From this relation, a fact can be found that the heavier the quark, the smaller the parameter $\mu_{ij}$. When extending to the heavy-quark case, there is no experimental data available, so we take it as a adjustable parameter. In Ref.~\cite{Huang:2015uda}, we investigate the mass spectrum of $P_{\psi}^N$ with $\mu_{cc}$ varying from $10^{-4}$ to $10^{-2}$ fm$^{-2}$, and our estimation indicated that the dependence of the parameter $\mu_{cc}$ is not very significant. In the present work, we take $\mu_{cc}=0.01\ \mathrm{fm}^{-2}$. Then $\mu_{sc}$ and $\mu_{nc}$ are obtained by the relations $\mu_{sc}^{2}=\mu_{ss}\mu_{cc} $ and $\mu_{nc}^{2}=\mu_{nn}\mu_{cc}$, respectively. It should be noted that $\mu_{ij}$ are phenomenal model parameters, their values are determined by reproducing the relevant mass spectra and phase shifts of the scattering processes. In Ref.~\cite{Chen:2011zzb}, the authors found that with the relation $\mu_{qs}^2=\mu_{qq} \mu_{ss}$, the masses of the ground state baryons composed of light quarks could be well reproduced. Later on, such relations have been successfully applied to investigate the states with heavy quarks~\cite{Huang:2013rla, Huang:2015uda,Liu:2022vyy}.


The Goldstone-boson exchange interactions between light quarks appear because of the dynamical breaking of chiral symmetry. The following $\pi$, $K$, and $\eta$ exchange terms work between the chiral quark-(anti)quark pair, which read,
\begin{eqnarray}
V_{\chi}(r_{ij}) & =&  v^{\pi}_{ij}(r_{ij})\sum_{a=1}^{3}\lambda_{i}^{a}\lambda_{j}^{a}+v^{K}_{ij}(r_{ij})\sum_{a=4}^{7}\lambda_{i}^{a}\lambda_{j}^{a}+v^{\eta}_{ij}(r_{ij})\nonumber\\
&&\left[\left(\lambda _{i}^{8}\cdot
\lambda _{j}^{8}\right)\cos\theta_P-\left(\lambda _{i}^{0}\cdot
\lambda_{j}^{0}\right) \sin\theta_P\right], \label{sala-Vchi1}
\end{eqnarray}
with
\begin{eqnarray}
  v^{B}_{ij} &=&  {\frac{g_{ch}^{2}}{{4\pi}}}{\frac{m_{B}^{2}}{{\
12m_{i}m_{j}}}}{\frac{\Lambda _{B}^{2}}{{\Lambda _{B}^{2}-m_{B}^{2}}}}
m_{B}     \nonumber    \\
&&\times\left\{(\boldsymbol{\sigma}_{i}\cdot\boldsymbol{\sigma}_{j})
\left[ Y(m_{B}\,r_{ij})-{\frac{\Lambda_{B}^{3}}{m_{B}^{3}}}
Y(\Lambda _{B}\,r_{ij})\right] \right\},
\end{eqnarray}
with $B=(\pi, K,  \eta)$ and $Y(x)=e^{-x}/x$ to be the standard Yukawa function. $\boldsymbol{\lambda^{a}}$ is the SU(3) flavor Gell-Mann matrix. The masses of the $\eta$, $K$, and $\pi$ meson are taken from the experimental value~\cite{ParticleDataGroup:2018ovx}. By matching the pion exchange diagram of the $NN$ elastic scattering process at the quark level and at the hadron level, one can relate the $\pi qq$ coupling with the one of $\pi NN$, which is~\cite{Vijande:2004he,Fernandez:1986zn},
\begin{equation}
\frac{g_{ch}^{2}}{4\pi}=\left(\frac{3}{5}\right)^{2} \frac{g_{\pi NN}^{2}}{4\pi} {\frac{m_{u,d}^{2}}{m_{N}^{2}}},
\end{equation}
which assumes that the flavor SU(3) is an exact symmetry, and only broken by the masses of the strange quark. As for the coupling $g_{\pi NN}$, it was determined by the $NN$ elastic scattering~\cite{Fernandez:1986zn}. All model parameters, except for those related to the strange quark flavor, are taken from Ref.~\cite{Liu:2023huu}, which were determined by reproducing the mass spectrum of the ground states mesons and baryons in QDCSM. Parameters for the strange quark are based on Ref.~\cite{Liu:2023oyc}. Detailed descriptions are omitted here for brevity. With those model parameters, the nature of the anticharmed-strange pentaquark states will be investigated in
QDCSM.

Besides, in QDCSM, the quark delocalization is realized by specifying the single-particle orbital wave function as a linear combination of left and right Gaussian basis, the single-
particle orbital wave functions used in the ordinary quark cluster model reads,
\begin{eqnarray}\label{wave0}
\psi_{\alpha}(\boldsymbol{s}_{i},\epsilon)&=&\left(\Phi_{\alpha}(\boldsymbol{s}_{i})
  +\epsilon\Phi_{\beta}(\boldsymbol{s}_{i})\right)/N(\epsilon), \nonumber \\
\psi_{\beta}(\boldsymbol{s}_{i},\epsilon)&=&\left(\Phi_{\beta}(\boldsymbol{s}_{i})
  +\epsilon\Phi_{\alpha}(\boldsymbol{s}_{i})\right)/N(\epsilon), \nonumber \\
N(\epsilon)&=& \sqrt{1+\epsilon^2+2\epsilon e^{-s^2_{i}/{4b^2}}},\nonumber \\
\Phi_{\alpha}(\boldsymbol{s}_{i})&=&\left(\frac{1}{\pi b^2}\right)^{\frac{3}{4}}
e^{-\frac{1}{2b^2}\left(\boldsymbol{r_\alpha}-\frac{2}{5}s_{i}\right)^2},\nonumber \\
\Phi_{\beta}(-\boldsymbol{s}_{i})&=&\left(\frac{1}{\pi b^2}\right)^{\frac{3}{4}}
e^{-\frac{1}{2b^2}\left(\boldsymbol{r_\beta}+\frac{3}{5}s_{i}\right)^2},
\end{eqnarray}
with $\boldsymbol{s}_{i}$, $i=(1,2,..., n)$, to be the generating coordinates, which are introduced to
expand the relative motion wave function~\cite{Wu:1998wu,Ping:1998si,Pang:2001xx}. The parameter $b$ indicates the size of the  baryon and meson clusters, which is determined by fitting the radius of the baryon and meson by the variational method~\cite{Huang:2018rpb}. In addition, The mixing parameter $\epsilon(s_{i})$ is not an adjusted one but is determined variationally by the dynamics of the the multi-quark system itself. This assumption allows the multi-quark system to choose its favorable configuration in the interacting process. It has been used to explain the cross-over transition between the hadron phase\footnote{The phase shift of $NN$ interaction could be described with the formalisms with hadrons only. After including the pseudo-scalar, vector and scalar meson, especially the $\sigma$ meson, the $NN$ interaction has been well described. In Ref.~\cite{Chen:2007qn}, the authors concluded that the $\sigma$-meson exchange can be replaced by quark delocalization and color screening mechanism introduced by QDCSM by comparing the NN scattering and deuteron properties obtained by chiral quark model and QDCSM} and the quark-gluon plasma phase~\cite{Xu:2007oam, Chen:2007qn, Huang:2011kf}. Due to the effect of the mixing parameter $\epsilon(s_{i})$, there is a certain probability for the quarks between the two clusters to run, which leads to the existence of color octet states for the two clusters. Therefore, this model also includes the hidden color channel effect, which is confirmed by Refs.~\cite{Xia:2021tof,Huang:2020bmb}.

\begin{table}[htb]
\begin{center}
\caption{\label{channels} The relevant channels for all possible states with different $J^P$ quantum numbers}
\renewcommand\arraystretch{1.5}
\begin{tabular}{p{1.0cm}<\centering p{0.8cm}<\centering p{0.8cm}<\centering p{0.8cm}<\centering p{0.8cm}<\centering p{0.8cm}<\centering p{0.8cm}<\centering p{0.8cm}<\centering p{0.8cm}<\centering p{1.0cm}<\centering p{1.0cm}<\centering p{1.0cm}<\centering p{1.0cm}<\centering p{1.0cm}<\centering p{1.0cm}<\centering p{1.0cm}<\centering p{1.0cm}<\centering p{1.0cm}<\centering}
\toprule[1pt]
  &    \multicolumn{2}{c}{$S=\frac{1}{2}$}  && \multicolumn{2}{c}{$S=\frac{3}{2}$} && \multicolumn{2}{c}{$S=\frac{5}{2}$}\\
\midrule[1pt]
\multirow{4}{*}{$I=\frac{1}{2}$} & $\Lambda D$ & $\Lambda D^{\ast}$ &&$\Lambda D^{\ast}$ &$\Sigma D^{\ast}$ &&$\Sigma^{\ast}D^{\ast}$ \\
& $\Sigma D$ & $\Sigma D^{\ast}$ &&$\Sigma^{\ast} D$ &$\Sigma^{\ast}D^{\ast}$\\
& $\Sigma^{\ast}D^{\ast}$&$ND_{s}$ && $ND_{s}^{\ast}$\\
&$N D_{s}^{\ast}$ \\
\midrule[1pt]
\multirow{3}{*}{$I=\frac{3}{2}$}& $\Sigma D$ &$\Sigma D^{\ast}$ && $\Sigma D^{\ast}$ & $\Sigma^{\ast}D$&& $\Sigma^{\ast}D^{\ast}$ &$\Delta D_{s}^{\ast}$\\
                                &$\Sigma^{\ast}D^{\ast}$ &$\Delta D_{s}^{\ast}$ && $\Sigma^{\ast}D^{\ast}$ &$\Delta D_{s}$\\
                                &&&& $\Delta D_{s}^{\ast}$\\
\bottomrule[1pt]
\end{tabular}
\end{center}
\end{table}

\section{The results and discussions\label{dis}}
In this work, we perform a comprehensive investigation of the low-lying anticharmed-strange pentaquarks with configuration $\bar{c}snnn$ within the QDCSM.  Here, we only consider the $S-$wave channels with the quantum numbers $I=1/2,3/2$ and $S=1/2,3/2,5/2$. According to the symmetry requirements of the four degrees of freedom, the possible physical channels under different numbers are listed in Table~\ref{channels}. In the present work, our purpose is to explore if there is any other pentaquark states and to see whether those pentaquark states can be explained as the molecular pentaquarks. Moreover, to better understand the properties of the pentauqark states, it is essential to calculate the magnetic moments of different pentaquark states.  Therefore, to achieve these objectives, the resonance group method  (RGM) and generator coordinates method are employed. The details of these methods can be found in Refs.~\cite{Liu:2023oyc,Kamimura:1977okl,wheeler1937molecular,Hill:1952jb,Griffin:1957zza}

 For the $c\bar{s}nnn$ pentaquark system, we only consider the $S-$wave channels with the spin $S=1/2, 3/2$ and $5/2$, respectively. This work mainly focuses on the pentaquark states in the molecular scenario, where pentaquark states are composed of a baryon cluster and a meson cluster.
 Considering the color structure of these two clusters, both the color singlet-singlet $(1_c\otimes 1_c)$ and the color octet-octet $(8_c \otimes 8_c)$ channels should be involved in principle. However, as noted in Ref. \cite{Huang:2011kf}, the  color screening confinement in Eq. (\ref{Eq:fr}) could be considered an effective description of the the color octet-octet channels (also known as the hidden color channels). Thus,  only the color singlet-singlet channels are taken into consideration in the present estimations, and all the possible channels involved are collected in Table~\ref{channels}.

\begin{figure}[t]
\includegraphics[scale=0.65]{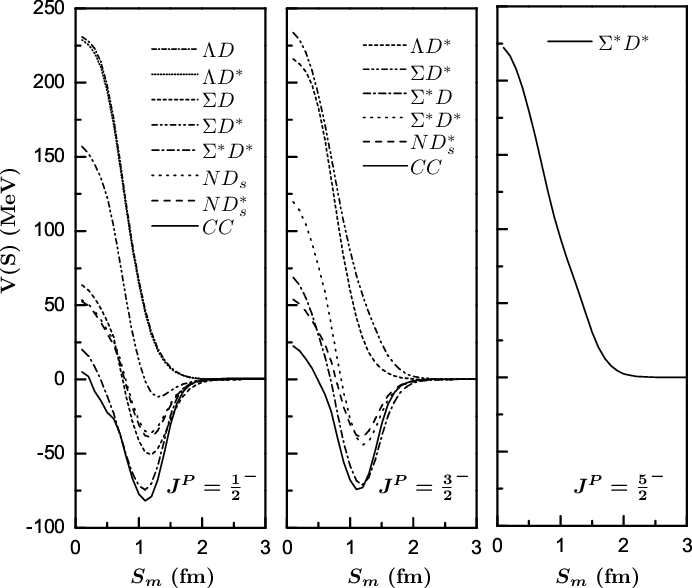}
 \caption{ The effective potentials defined in Eq.~(\ref{Eq:PotentialV}) for different channels of the anticharmed-strange pentaquark systems with $I=1/2$ in QDCSM.$CC$ stands for the effective potentials after considering the all channel coupling. }
\label{Veff-0.5}
\end{figure}

\begin{figure}[t]
\includegraphics[scale=0.65]{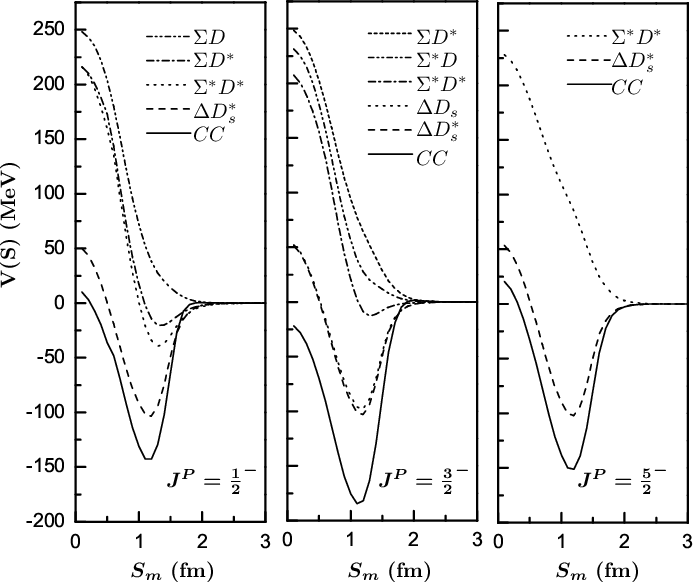}
 \caption{ The effective potentials defined in Eq.~(\ref{Eq:PotentialV}) for different channels of the anticharmed-strange pentaquark systems with $I=3/2$ in QDCSM. $CC$ stands for the effective potentials after considering the all channel coupling }
\label{Veff-1.5}
\end{figure}

\subsection{The effective potentials}
To search the possible bound states and resonance states composed of the hadron pair listed in Table~\ref{channels}, we first estimate the effective potentials between these hadron pairs. The potential is defined as
\begin{eqnarray}
E(S_{m})&=&\frac{\langle\Psi_{5q}(S_m)|H|\Psi_{5q}(S_m)\rangle}{\langle\Psi_{5q}(S_m)|\Psi_{5q}(S_m)\rangle},\label{Eq:PotentialE}
\end{eqnarray}
where $S_m$ denotes the distance between two clusters and $\Psi_{5q}(S_m)$ represents the wave function of a certain given channel. The terms  $\langle\Psi_{5q}(S_m)|H|\Psi_{5q}(S_m)\rangle$ and $\langle\Psi_{5q}(S_m)|\Psi_{5q}(S_m)\rangle$ correspond to the Hamiltonian matrix and the overlap of the states, respectively. Thus, the effective potential between two colorless cluster is defined as,
\begin{eqnarray}
	V(S_m)=E(S_m)-E(\infty), \label{Eq:PotentialV}
\end{eqnarray}
where $E(\infty)$ represents the energy at a sufficient large distance of two clusters. The estimated potentials for $I=1/2$ and $I=3/2$ are presented in Fig.~\ref{Veff-0.5} and \ref{Veff-1.5}, respectively.

From the Fig.~\ref{Veff-0.5}, the seven single channels with $I(J^{P})=\frac{1}{2}(\frac{1}{2}^{-})$ are identified as $\Lambda D$, $\Lambda D^{\ast}$, $\Sigma D$, $\Sigma D^{\ast}$, $\Sigma^{\ast} D^{\ast}$, $N D_{s}$ and $N D_{s}^{\ast}$, respectively. The effective potentials for the $\Lambda D$ and $\Lambda D^{\ast}$ are repulsive whereas the remaining channels exhibit attractive potentials. Notably, the effective attraction between $\Sigma^{\ast} D^{\ast}$ and $\Sigma D$ is stronger than that of $N D_{s}^{\ast}$, $N D_{s}$, and $\Sigma D^{\ast}$, suggesting that $\Sigma^{\ast} D^{\ast}$ and $\Sigma D$ are likely to form bound or resonance states. Furthemore, Fig.~\ref{Veff-0.5} presents the effective potentials after channel coupling. The strong attraction observed after channel coupling indicates a high likelihood of forming bound states.
For the case of $I(J^{P})=\frac{1}{2}(\frac{3}{2}^{-})$, the effective potentials of the $\Lambda D^{\ast}$ and $\Sigma D^{\ast}$ are repulsive, thus preventing the formation of bound states. Among the remaining attractive channels, the effective attraction between $\Sigma^{\ast}$ and $D$ is significantly stronger compared to other channels, suggesting that the $\Sigma^{\ast} D$ channel is highly likely to form a bound state. Channel coupling further deepens this attraction from Fig.~\ref{Veff-0.5}, increasing the likelihood of forming a bound state with $I(J^{P})=\frac{1}{2}(\frac{3}{2}^{-})$. For the $I(J^{P})=\frac{1}{2}(\frac{5}{2}^{-})$ case, only the $\Sigma^{\ast} D^{\ast}$ channel exists due to the S-wave limitation. The repulsive effective potential of the $\Sigma^{\ast} D^{\ast}$ channel precludes the possibility of forming a bound state.

Fig.~\ref{Veff-1.5} shows the effecive potentials for each channel with $I=3/2$. For $J^{P}=\frac{1}{2}^{-}$, it can be seen that the channels $\Delta D_{s}^{\ast}$, $\Sigma^{\ast}D^{\ast}$ and $\Sigma D^{\ast}$ have attractive potentials, except for the $\Sigma D$ channel. This indicates that bound states or resonance states are possible for the $\Delta D_{s}^{\ast}$, $\Sigma^{\ast}D^{\ast}$ and $\Sigma D^{\ast}$ channles, whereas no such states are expected for the $\Delta D_{s}^{\ast}$, $\Sigma^{\ast}D^{\ast}$ and $\Sigma D^{\ast}$ channel. Additionally, the attraction between $\Delta$ and  $D_{s}^{\ast}$  is the strongest, followed by that of the $\Sigma^{\ast}D^{\ast}$, which is slightly stronger than that of $\Sigma D^{\ast}$. The enhanced attractive potential following channel coupling implies the possibility of obtaining a bound state in the channel coupling estimation. For $J^{P}=\frac{3}{2}^{-}$, the potentials of the $\Delta D_{s}^{\ast}$, $\Delta D_{s}$ and $\Sigma^{\ast} D^{\ast}$ channels are attractive, whereas the other two channels exhibit repulsive potentials. The attractions in the $\Delta D_{s}^{\ast}$ and $\Delta D_{s}$ channels, which are nearly the same, are much stronger than in the $\Sigma^{\ast} D^{\ast}$ channel, indicating the potential for bound states in the $\Delta D_{s}^{\ast}$ and $\Delta D_{s}$ channels. After channel coupling, the strongest attraction suggests a high probability of a bound state. For the case of $J^{P}=\frac{5}{2}^{-}$, there are only two channels, which are $\Sigma^{\ast} D^{\ast}$ and $\Delta D_{s}^{\ast}$, respectively. From Fig.~\ref{Veff-1.5}, the potential for the $\Delta D_{s}^{\ast}$ channel is attractive, while the potential for the $\Sigma^{\ast} D^{\ast}$  channel is repulsive. Nevertheless, after channel coupling, the strongest attractive potential suggests the possibility of obtaining a bound state.

\begin{table*}[htb]
\begin{center}
\renewcommand{\arraystretch}{1.5}
\caption{\label{bound-1} The binding energies and the masses of every single channel and those of channel coupling for the pentaquarks with $I=1/2$. The values are provided in units of MeV. }
\begin{tabular}{p{1.6cm}<\centering p{1.6cm}<\centering p{1.6cm}<\centering p{1.6cm}<\centering p{1.6cm}<\centering p{1.6cm}<\centering p{1.6cm}<\centering p{1.6cm}<\centering p{2.0cm}<\centering p{1.6cm}<\centering  }
\toprule[1pt]
$I(J^{P})$  & Channel & $E_{sc}$ &$E_{th}^{Model}$ &$E_{B}$ &$E_{th}^{Exp}$ &$E_{sc}^{\prime}$ & $E_{cc}/E_B$ & $E_{cc}^\prime$ \\
 \midrule[1pt]
\multirow{7}{*}{$\frac{1}{2}(\frac{1}{2}^{-})$} &$\Lambda D$                &2992 &2988 &4 &3122 &3126 &\multirow{7}{*}{2873/-21}&\multirow{7}{*}{2886}\\
                                                &$\Lambda D^{\ast}$         &3027 &3023 &4 &3227 &3231\\
                                                &$\Sigma D$                 &3101 &3102 &-1&3058 &3057\\
                                                &$\Sigma D^{\ast}$          &3140 &3137 &3 &3196 &3199\\
                                                &$\Sigma^{\ast} D^{\ast}$   &3251 &3259 &-8&3392 &3384\\
                                                &$N D_{s}$                  &2896 &2894 &2 &2907 &2909\\
                                                &$N D_{s}^{\ast}$           &2910 &2908 &2 &3051 &3053\\
\midrule[1pt]
\multirow{5}{*}{$\frac{1}{2}(\frac{3}{2}^{-})$} &$\Lambda D^{\ast}$         &3027 &3023 &4 &3122 &3126 &\multirow{5}{*}{2896/-12}&\multirow{5}{*}{3039}\\
                                                &$\Sigma D^{\ast}$          &3141 &3137 &4 &3196 &3200\\
                                                &$\Sigma^{\ast} D$          &3218 &3224 &-6&3254 &3248\\
                                                &$\Sigma^{\ast} D^{\ast}$   &3261 &3259 &2 &3392 &3394\\
                                                &$N D_{s}^{\ast}$           &2910 &2908 &2 &3051 &3053\\
\midrule[1pt]
$\frac{1}{2}(\frac{5}{2}^{-})$                  &$\Sigma^{\ast} D^{\ast}$   &3263 &3259 &4 &3392 &3396 &3263/4 &3396\\

\bottomrule[1pt]
\end{tabular}
\end{center}
\end{table*}

\begin{table*}[htb]
\begin{center}
\renewcommand{\arraystretch}{1.5}
\caption{\label{bound-2} The binding energies and the masses of every single channel and those of channel coupling for the pentaquarks with $I=3/2$. The values are provided in units of MeV.}
\begin{tabular}{p{1.6cm}<\centering p{1.6cm}<\centering p{1.6cm}<\centering p{1.6cm}<\centering p{1.6cm}<\centering p{1.6cm}<\centering p{1.6cm}<\centering p{1.6cm}<\centering p{2.0cm}<\centering p{1.6cm}<\centering  }
\toprule[1pt]
$I(J^{P})$  & Channel & $E_{sc}$ &$E_{th}^{Model}$ &$E_{B}$ &$E_{th}^{Exp}$ &$E_{sc}^{\prime}$ &$E_{cc}/E_B$& $E_{cc}^\prime$\\
 \midrule[1pt]
\multirow{4}{*}{$\frac{3}{2}(\frac{1}{2}^{-})$}   &$\Sigma D$                 &3106 &3102 &4   &3058 &3062  &\multirow{4}{*}{3103/1}&\multirow{4}{*}{3059}\\
                                                  &$\Sigma D^{\ast}$          &3140 &3137 &3   &3196 &3199\\
                                                  &$\Sigma^{\ast} D^{\ast}$   &3261 &3259 &2   &3392 &3394\\
                                                  &$\Delta D_{s}^{\ast}$      &3167 &3201 &-34 &3344 &3310\\
\midrule[1pt]
\multirow{5}{*}{$\frac{3}{2}(\frac{3}{2}^{-})$}   &$\Sigma D^{\ast}$          &3141 &3137 &4   &3196 &3200
&\multirow{5}{*}{3094/-43}&\multirow{5}{*}{3153}\\
                                                  &$\Sigma^{\ast} D$          &3228 &3224 &4   &3254 &3248\\
                                                  &$\Sigma^{\ast} D^{\ast}$   &3263 &3259 &4   &3392 &3394\\
                                                  &$\Delta D_{s}$             &3162 &3187 &-25 &3200 &3175\\
                                                  &$\Delta D_{s}^{\ast}$      &3155 &3201 &-45 &3344 &3299\\
\midrule[1pt]
\multirow{2}{*}{$\frac{3}{2}(\frac{5}{2}^{-})$}   &$\Sigma^{\ast} D^{\ast}$   &3263 &3259 &4   &3392 &3396
&\multirow{2}{*}{3202/1}&\multirow{2}{*}{3345}\\
                                                  &$\Delta D_{s}^{\ast}$      &3206 &3201 &5   &3344 &3349\\
\bottomrule[1pt]
\end{tabular}
\end{center}
\end{table*}

\subsection{Possible bound states}
Given the potential analysis, dynamical calculations are essential to determine the existence of bound states or resonance states in the $\bar{c}snnn$ system. It should be noted in single channel calculations, strong attraction between two involved hadrons makes the formation of bound states or resonance states highly probable. Additionally, in multi-channel coupling calculations, if the obtained lowest energy is below the theoretical threshold of the lowest channel, it indicates the presence of a bound state, as seen the deuteron in two channel coupling calculation of $NN$ and $\Delta\Delta$ with $I(J^{P})=0(1^{+})$~\cite{Huang:2011kf}. Moreover, it is important to note that the present calculations are conducted within a finite space, indicating a limited number of basis functions. Therefore, even for repusive potentials in single channel estmations, a series of basis can be obtained. However, the obtained energy in this case should not be the eigenenergy of a bound state since the repusive potential cannot form a bound state. These obtained eigenenergies are higher than the corresponding channel thresholds and mainly depend on the number of basis. When the space is sufficiently large, the eigenenergies will approach the corresponding threshold. For this case, the corresponding eigenenergies are listed in Tables~\ref{bound-1} and \ref{bound-2} to figure out the channel coupling effect in the present estimations. For the bound states, the estimated eigenvalues are below the threshold and remian quite stable as the space increases, which has been further validated in our estimations. Overall, the lowest eigenenergies of the identified bound states must fall below the theoretical threshold of the corresponding lowest channel and be stable against strong decay.

The estimated results are detailed in Tables ~\ref{bound-1} and \ref{bound-2}, corresponding to states with $I=1/2$ and $I=3/2$, respectively. In these tables, $E_{sc}$ and $E_{cc}$ denote the eigenenergies of the $\bar{c}snnn$ pentaquark states from single channel and channel coupling estimations. $E_{th}^{Model}$ and $E_{th}^{Exp}$ represent the theoretical estimations and experimental measurements of the channel thresholds, respectively. The binding energy, $E_{B}=E_{sc}-E_{th}^{Model}$, is derived from single channel estimations. It is important to note that in this work, the relevant parameters are determined based on various aspects of hadron properties. The inaccuracy of model parameters will lead to uncertainty in the model predictions.
However, compared to the absolute values of the eigenenergies, the mass splittings, such as the $E_B$, are more reliable. Furthermore, in the present estimations, we define the corrected eigenenergy of the single channel estimations $E_{sc}^\prime$ as $E_{th}^{Exp}+E_B$. Similarly, by taking the lowest threshold of the involved channels as a reference, we can obtain the corrected eigenenergy of the coupled-channel estimations  $E_{cc}^\prime$. This approach helps reduce the model dependence of the corrected eigenenergies to some extent.

For the system with $I(J^{P})=\frac{1}{2}(\frac{1}{2}^{-})$,  Table~\ref{bound-1} shows that $\Sigma^{\ast} D^{\ast}$ and $\Sigma D$ states in single channel estimations can be bound states with the binding energies to be $-8$ MeV and $-1$ MeV, respectively. In relation to Fig.~\ref{Veff-0.5}, it can be seen that the strong attraction of $\Sigma^{\ast} D^{\ast}$ and $\Sigma D$ states in single channel estimations leads to eigenvalues for these two states that fall below the theoretical thresholds of their respective channels. However, for the $ND_{s}^{\ast}$, $ND_{s}$ and $\Sigma D^{\ast}$ states, the eigenenergies obtained in single channel estimations are all above the theoretical thresholds of the corresponding channels due to the rather weak attraction, indicating that these states become scattering states.  In the case of $\Lambda D$ and $\Lambda D^{\ast}$, the extreme repulsion for these two states leads to the obtained eigenengies that exceed the threshold of $\Lambda D$ and $\Lambda D^{\ast}$, so these two states also become scattering states. For the channel coupling estimations, the effective potential of channel coupling, as shown in Fig.~\ref{Veff-0.5}, indicates strong attraction between the two hadrons. Therefore, the channel coupling estimation shows that a bound state with the binding energy to be -21 MeV can be formed, with a corrected mass and rms of 2886 MeV and 1.2 fm, respectively. Additionally, to determine the composition of the bound state derived from the channel coupling calculations, we perform relevant calculations. The outcome reveals that component $N D_{s}$ state comprises $96\%$ of the bound state.

Based on the above analysis, the effective attraction for $\Sigma^{\ast} D^{\ast}$ and $\Sigma D$ states results in their transformation into bound states in single channel estimations, with the obtained eigenengies exceeding the lowest threshold. This indicates that these bound states can couple with the corresponding channels and decay through those channels. With the inclusion of channel coupling effects, these bound states are likely to transform into resonance states in certain scattering processes. To verify this possibility, we will undertake relevant estimations to locate resonance states in the scattering processes of specific open channels in the next subsection.

For the system with $I(J^{P})=\frac{1}{2}(\frac{3}{2}^{-})$, a bound state $\Sigma^{\ast} D$ with binding energy of -6 MeV can be obtained due to the stronger attraction between $\Sigma^{\ast}$ and $D$ in single channel estimations. However, the eigenenergies of the other channels are above the theoretical thresholds, indicating that they are scattering states. From Fig.~\ref{Veff-0.5}, although the effective potentials for $\Sigma^{\ast}D^{\ast}$ and $N D_{s}^{\ast}$ are attractive, the weak attraction prevents the formation of bound states. Conversely, $\Lambda D^{\ast}$ and $\Sigma D^{\ast}$ cannot form bound states due to repulsion. Through muti-channel coupling estimations, a bound state with a binding energy of -12 MeV and a corrected mass of 3039 MeV is found. Composition and distance calculations indicate that the distance between the two clusters is about 1.5 fm and the bound state is predominantly composed of $ND_{s}^{\ast}$.  For the system with $I(J^{P})=\frac{1}{2}(\frac{5}{2}^{-})$, there is only a channel $\Sigma^{\ast} D^{\ast}$, the obtained eigenenergies for $\Sigma^{\ast} D^{\ast}$ is above the theoretical threshold of $\Sigma^{\ast} D^{\ast}$ due to the repulsive interaction between $\Sigma^{\ast}$ and $ D^{\ast}$, so $\Sigma^{\ast} D^{\ast}$ state cannot form a bound state.

For the system with $I(J^{P})=\frac{3}{2}(\frac{1}{2}^{-})$, There are four channels: $\Sigma D$, $\Sigma D^{\ast}$, $\Sigma^{\ast} D^{\ast}$, and $\Delta D_{s}^{\ast}$, respectively. The single channel estimations indicate that the strong attraction between $\Delta $ and $D_{s}^{\ast}$ can form a bound state $\Delta D_{s}^{\ast}$ with a binding energy of -34 MeV. In contrast, the weak attraction in the $\Sigma^{\ast} D^{\ast}$ and $\Sigma D^{\ast}$ channels is insufficient to form bound states. For the $\Sigma D$ state, the lowest eigenenergy obtained is higher than the threshold of physical channel $\Sigma D$ due to repulsive interaction in single channel estimations. This indicates that the $\Sigma D$ state, being a scattering state, cannot form a bound state.
Although the effective potential of channel coupling exhibits strong attraction, the lowest eigenenergy obtained is above the threshold of the lowest physical channel by 1 MeV.

For the system with $I(J^{P})=\frac{3}{2}(\frac{3}{2}^{-})$, the $\Delta D_{s}^{\ast}$ and $\Delta D_{s}$ states are bound states in single channel estimations, with binding energies of -45 MeV and -25 MeV, respectively. These results align with the expected behavior of $\Delta D_{s}^{\ast}$ and $\Delta D_{s}$ states in single channel estimation. These two bound states can decay into open channels, and the nature of these bound states will be further examined in the scattering of these open channels, as presented in the next subsection. Additionally, single channel estimations for the remaining three physical channels reveal that they are scattering states, with eigenenergies surpassing their respective thresholds. However, when the channel coupling effect is considered, a bound state with a mass of approximately 3153 MeV emerges. Analysis of this bound state's composition and distance between the two clusters indicates it is primarily composed of $\Sigma D^{\ast}$, with an inter-cluster distance of 2.1 fm, pointing to a hadronic molecular state. For the system with $I(J^{P})=\frac{3}{2}(\frac{5}{2}^{-})$,  $\Sigma^{\ast}D^{\ast}$ and $\Delta D_{s}^{\ast}$ states are given. The present results from both single channel and channel coupling estimations show that no bound states exist when only S-wave interaction is considered.

\begin{figure}
\includegraphics[scale=0.65]{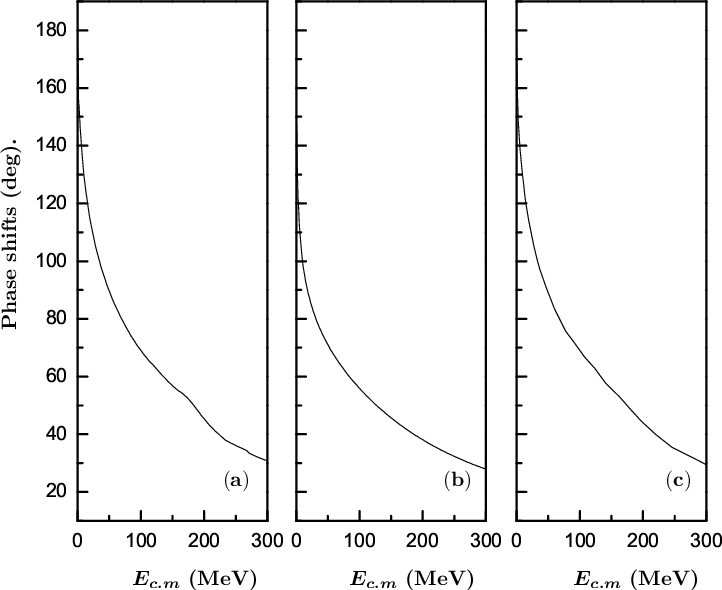}
 \caption{ The phase shifts of the $ND_{s}$ with $I(J^{P})=\frac{1}{2}(\frac{1}{2}^{-})$, $N D_{s}^{\ast}$ with $I(J^{P})=\frac{1}{2}(\frac{3}{2}^{-})$ and $\Sigma D^{\ast}$ with $I(J^{P})=\frac{3}{2}(\frac{3}{2}^{-})$ in the coupled channel estimations. (a) stands for $ND_{s}$ with $I(J^{P})=\frac{1}{2}(\frac{1}{2}^{-})$, (b) represents $N D_{s}^{\ast}$ with $I(J^{P})=\frac{1}{2}(\frac{3}{2}^{-})$, (c) stands for $\Sigma D^{\ast}$ with $I(J^{P})=\frac{3}{2}(\frac{3}{2}^{-})$.}
\label{low-phase}
\end{figure}

The dynamical calculations reveal three bound states in the $\bar{c}snnn$ pentaquark system under the quantum numbers $I(J^{P})=\frac{1}{2}(\frac{1}{2}^{-})$, $I(J^{P})=\frac{1}{2}(\frac{3}{2}^{-})$ and $I(J^{P})=\frac{3}{2}(\frac{3}{2}^{-})$. To further confirm these bound states,  we first utilize the well-developed Kohn-Hulthen-Kato (KHK) variational method to investigate their low-energy scattering behavior, as detailed in Ref.~\cite{Kamimura:1977okl}.  As shown in Fig~\ref{low-phase}, at near-zero incident energy, the scattering phase shift approaches 180 degrees and rapidly decreases as the incident energy increases, further supporting the existence of three bound states. We then derive physical parameters of the low-energy scattering phase, such as scattering length $a_{0}$, effective range $r_{0}$, and binding energy $E_{B}^{\prime}$ by using the formula:
\begin{eqnarray}\label{wave13}
    k \cot{\delta_L} &=& -\frac{1}{a_{0}}+\frac{1}{2}r_{0}k^{2}+O(k^4),
\end{eqnarray}
where $k=\sqrt{2\mu E_{c.m}}$, $\mu$ and $E_{c.m}$ are the reduced mass of two hadrons and the incident energy, respectively. $\delta_{L}$ is the low-energy phase shifts obtained above.

According to above results, the wave number $\alpha$ can be derived using the relation~\cite{Babenko:2003js},
\begin{eqnarray}\label{wave14}
    r_{0}&=&\frac{2}{\alpha}\left(1-\frac{1}{\alpha a_{0}} \right).
\end{eqnarray}
Finally, the binding energy $E_B^{\prime}$ is calculated according to the relation,
\begin{eqnarray}\label{wave15}
    E_{B}^{\prime}=\frac{\hbar^{2}\alpha^{2}}{2 \mu}.
\end{eqnarray}

We can also use the aforementioned method to estimate the binding energy, and the results obtained have been listed in Table~\ref{bound-3}. From the table, it can be seen that the scattering lengths of the three bound states are positive and their binding energies are consistent with those obtained from dynamical calculations, confirming the presence of these bound states in the $\bar{c}snnn$ pentaquark system.


\begin{table}[t]
\begin{center}
\renewcommand{\arraystretch}{1.5}
\caption{\label{bound-3} The the scattering length $a_{0}$, the effective range $r_{0}$ and the binding energy $E_{B}^{\prime}$ determined by the variation method. }
\begin{tabular}{p{1.5cm}<\centering p{1.5cm}<\centering p{1.5cm}<\centering p{1.5cm}<\centering p{1.5cm}<\centering   }
\toprule[1pt]
$I(J^{P})$  & Channel & $a_{0}$ (fm) &$r_{0}$ (fm) &$E_{B}^{\prime}$ (MeV) \\
\midrule[1pt]
$\frac{1}{2}(\frac{1}{2}^{-})$ & $N D_{s}$               &2.06 &0.98 & -23.8 \\
$\frac{1}{2}(\frac{3}{2}^{-})$ & $N D_{s}^{\ast}$        &2.00 &1.00 & -11.7\\
$\frac{3}{2}(\frac{3}{2}^{-})$ & $\Sigma D^{\ast}$       &2.56 &1.45 & -40.7\\
\bottomrule[1pt]
\end{tabular}
\end{center}
\end{table}

\subsection{Possible resonance states}
\begin{table*}[htb]
\begin{center}
\renewcommand{\arraystretch}{1.5}
\caption{\label{R} The masses and decay widths (in the unit of MeV) of resonance states with the difference scattering process.
$M_{R}^{th}$ stands for the sum of the corresponding theoretical threshold of the open channel and the incident energy, $M_{R}$ represents the modified resonance mass. $\Gamma_{i}$ is the partial decay width of the resonance state decaying to the $i-$th open channel. $\Gamma_{Total}$ is the total decay width of the resonance state.}
\begin{tabular}{p{2.3cm}<\centering p{1.cm}<\centering p{1.cm}<\centering p{0.5cm}<\centering p{0.01cm}<\centering p{1.cm}<\centering p{1.cm}<\centering p{0.5cm}<\centering p{0.01cm}<\centering p{1.cm}<\centering p{1.cm}<\centering p{0.5cm}<\centering p{0.01cm}<\centering p{1.cm}<\centering p{1.cm}<\centering p{0.5cm}<\centering p{0.01cm}<\centering p{1.cm}<\centering p{1.cm}<\centering p{1.cm}<\centering p{1.cm}<\centering p{1.cm}<\centering p{1.cm}<\centering   p{2.cm}<\centering  }
\toprule[1pt]
\multirow{4}{*}{Open channels} & \multicolumn{7}{c}{Three channel coupling} &  & \multicolumn{11}{c}{Two channel coupling} \\
\cline{2-8} \cline{10-20}
\multicolumn{1}{c}{}  &\multicolumn{7}{c}{$I(J^{P})=\frac{1}{2}(\frac{1}{2}^{-})$} &  &\multicolumn{7}{c}{$I(J^{P})=\frac{1}{2}(\frac{1}{2}^{-})$} & &\multicolumn{3}{c}{$I(J^{P})=\frac{1}{2}(\frac{3}{2}^{-})$} \\
 \cline{2-8} \cline{10-16} \cline{18-20}
   &\multicolumn{3}{c}{$\Sigma D$} &  &\multicolumn{3}{c}{$\Sigma^{\ast} D^{\ast}$} &  &\multicolumn{3}{c}{$\Sigma D$} &  &\multicolumn{3}{c}{$\Sigma^{\ast} D^{\ast}$} & &\multicolumn{3}{c}{$\Sigma^{\ast} D$} \\
   \cline{2-4}\cline{6-8}\cline{10-12}\cline{14-16}\cline{18-20}
                      & $M_{R}^{th}$ & $M_{R}$ & $\Gamma_{i}$  &     &$M_{R}^{th}$ & $M_{R}$ &$\Gamma_{i}$      & & $M_{R}^{th}$ & $M_{R}$ & $\Gamma_{i}$  &  & $M_{R}^{th}$ & $M_{R}$ & $\Gamma_{i}$ & & $M_{R}^{th}$ & $M_{R}$ &$\Gamma_{i}$ \\
 \midrule[1pt]
 $N D_{s}$            &3097  &3053     &5.5  &   &$\dots$   &$\dots$  &$\dots$ &  &3098 &3054     &5.6  &&$\dots$      &$\dots$         &$\dots$ &&$\dots$      &$\dots$         &$\dots$\\
 $N D_{s}^{\ast}$     &3098  &3054     &7.5  &   &$\dots$   &$\dots$  &$\dots$ &  &3099 &3055     &7.8  &&$\dots$      &$\dots$         &$\dots$ &&$\dots$      &$\dots$         &$\dots$\\
 $\Lambda D$          &$\dots$    &$\dots$  &$\dots$  &   &$\dots$   &$\dots$  &$\dots$ &  &$\dots$    &$\dots$  &$\dots$  &&3258 &3390     &1.1  &&$\dots$      &$\dots$         &$\dots$\\
 $\Lambda D^{\ast}$   &$\dots$    &$\dots$  &$\dots$  &   &$\dots$   &$\dots$  &$\dots$ &  &$\dots$    &$\dots$  &$\dots$  &&$\dots$    &$\dots$  &$\dots$  &&3220      &3250         &1.3\\
 $\Sigma D^{\ast}$     &$\dots$    &$\dots$  &$\dots$  &   &$\dots$   &$\dots$  &$\dots$ &  &$\dots$    &$\dots$  &$\dots$  &&3257 &3389     &9.5  &&3222      &3252         &3.1\\
 $\Gamma_{Total}$     &              &         &13.0      &     &     &         &  & &              &         &13.4      &     &     &         &10.4 & &              &         &4.4    \\
\bottomrule[1pt]
\end{tabular}
\end{center}
\end{table*}

\begin{figure*}
\includegraphics[scale=0.35]{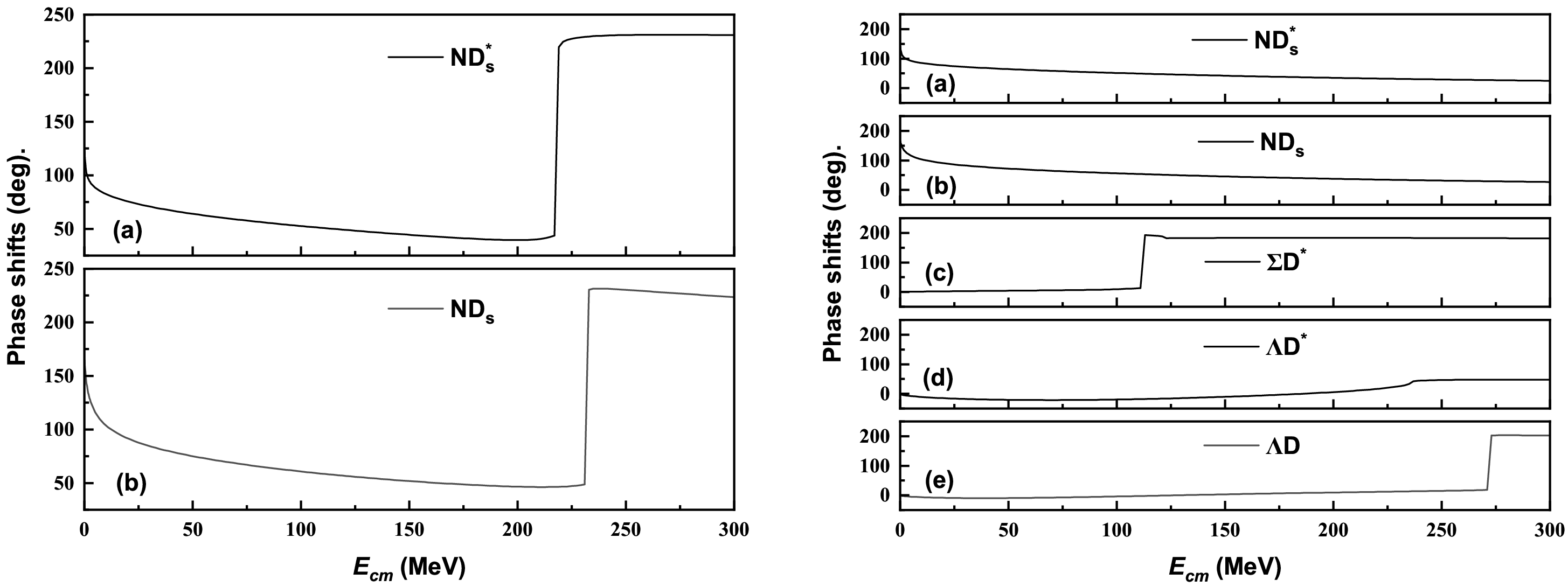}
 \caption{ The phase shifts of the open channels with two channel coupling for $I(J^{P})=1/2(1/2^{-})$ in QDCSM. On the left, (a) corresponds to two channels coupling with $\Sigma D$ and $N D_{s}^{\ast}$, (b) denotes two channel coupling with $\Sigma D$ and $N D_{s}$. On the right, (a) stands for two channel coupling with $\Sigma^{\ast} D^{\ast}$ and  $N D_{s}^{\ast}$, (b) indicates two channel coupling with  $\Sigma^{\ast} D^{\ast}$ and $N D_{s}$, (c) represents two channel coupling with  $\Sigma^{\ast} D^{\ast}$ and $\Sigma D^{\ast}$, (d) stands for two channel coupling with  $\Sigma^{\ast} D^{\ast}$ and $\Lambda D^{\ast}$, (e) denotes two channel coupling with  $\Sigma^{\ast} D^{\ast}$ and $\Lambda D$.}
\label{R-0.5-0.5-twocoupling}
\end{figure*}

\begin{figure}
\includegraphics[scale=0.35]{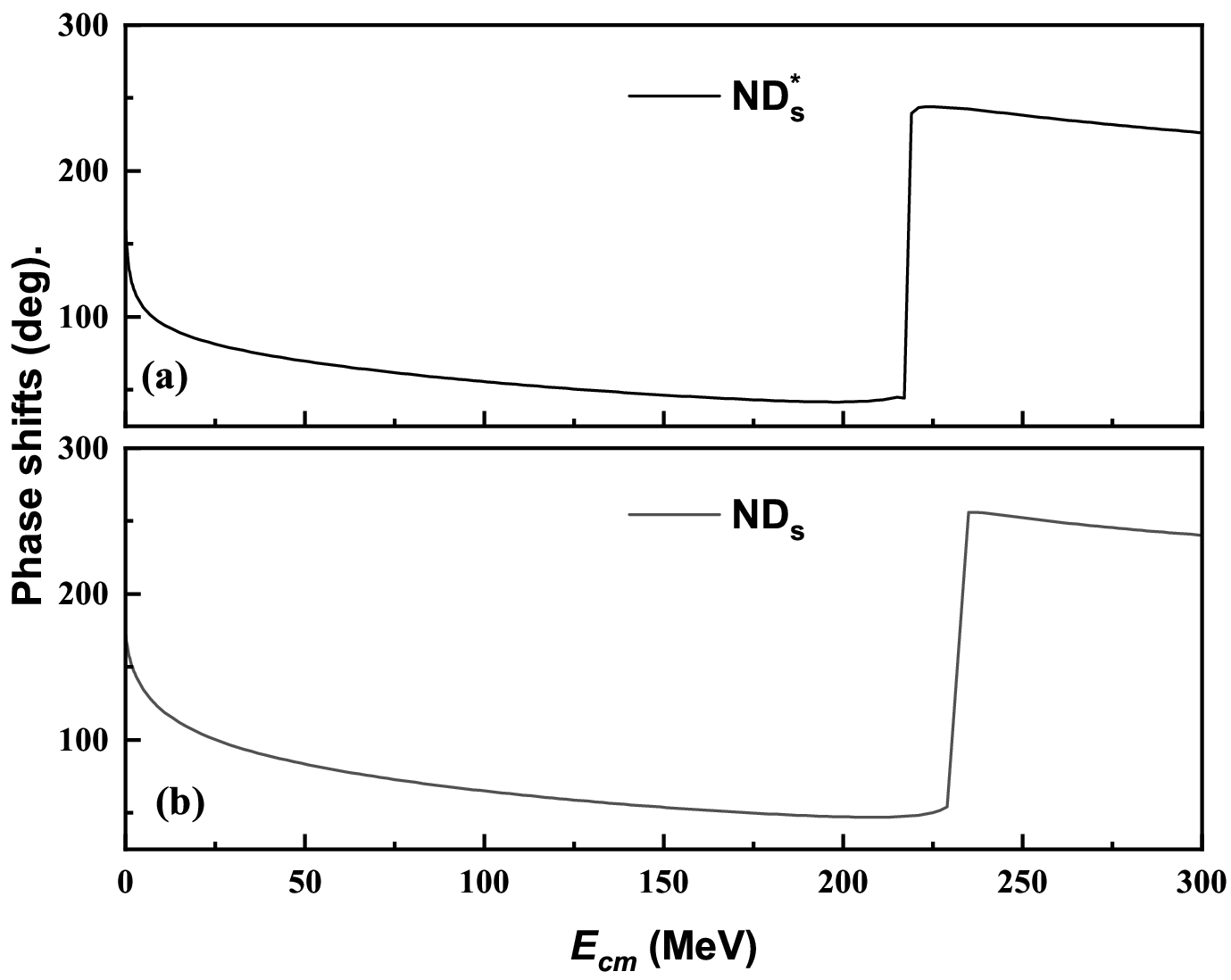}
 \caption{The phase shifts of the open channels with three channels coupling with two close channels ($\Sigma D$ and $\Sigma^{\ast} D^{\ast}$) and one open channel($N D_{s}$ or $N D_{s}^{\ast}$)for $I(J^{P})=1/2(1/2^{-})$ in QDCSM. }
\label{R-0.5-0.5-threecoupling}
\end{figure}

\begin{figure}
\includegraphics[scale=0.35]{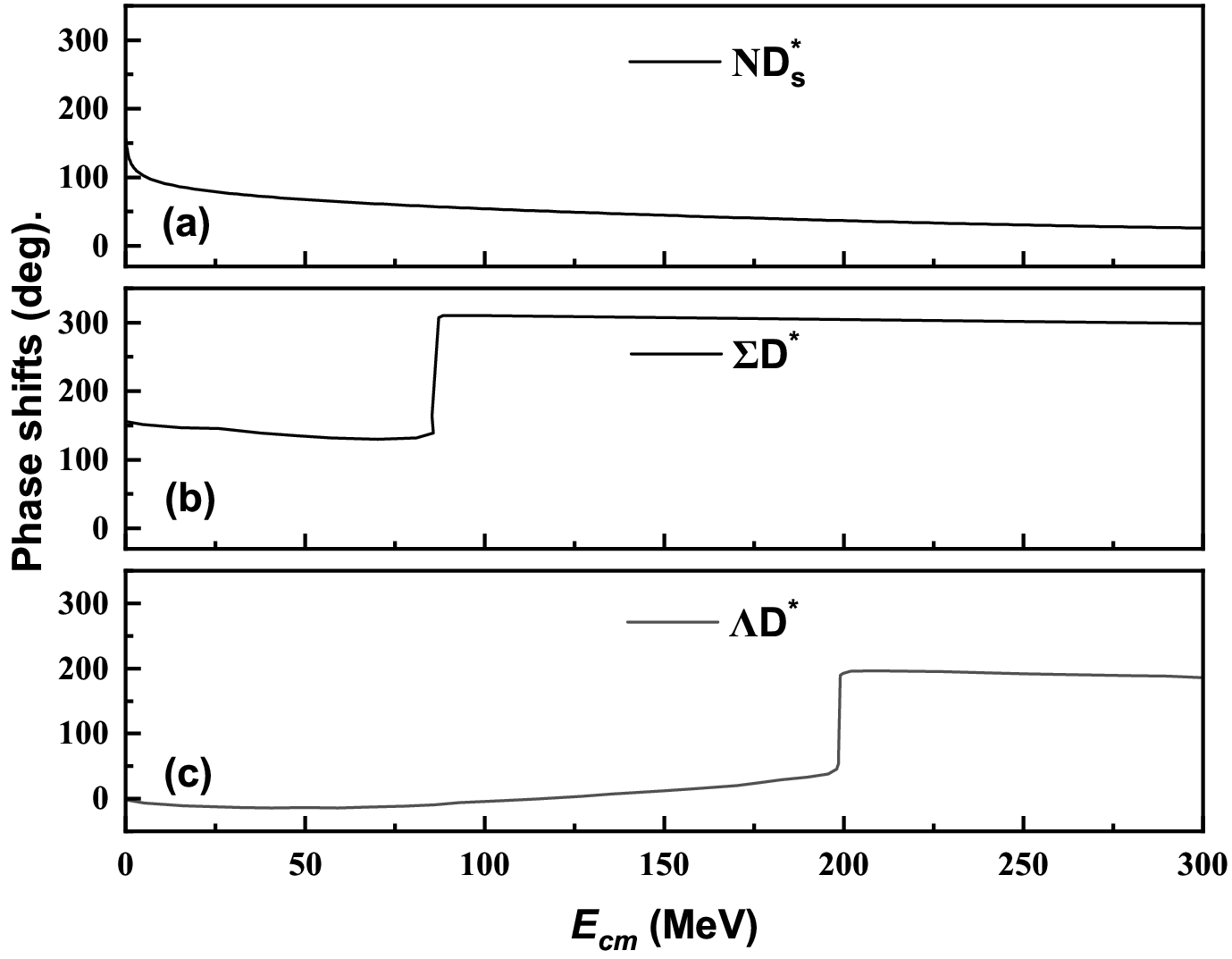}
 \caption{The phase shifts of the open channels with two channels coupling with a close channel($\Sigma^{\ast}D$) and one open channel($ND^{\ast}_{s}$, $\Sigma D^{\ast}$, $\Lambda D^{\ast}$) for $I(J^{P})=1/2(3/2^{-})$ in QDCSM. }
\label{R-0.5-1.5-twocoupling}
\end{figure}

\begin{figure}
\includegraphics[scale=0.35]{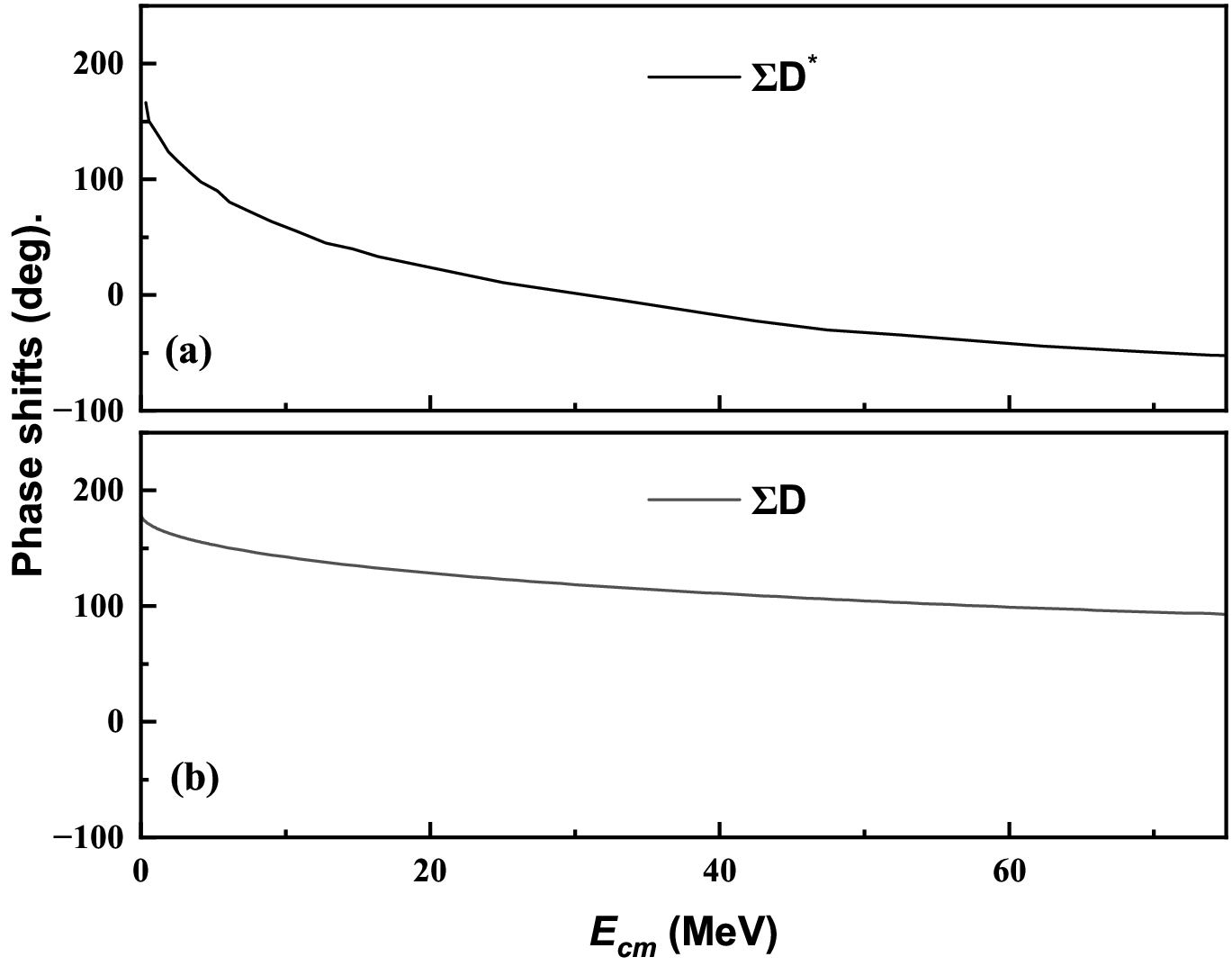}
 \caption{The phase shifts of the open channels with two channels coupling with a close channel($\Delta D_{s}^{\ast}$) and one open channel($\Sigma D$, $\Sigma D^{\ast}$) for $I(J^{P})=3/2(1/2^{-})$ in QDCSM. }
\label{R-1.5-0.5-twocoupling}
\end{figure}

\begin{figure}
\includegraphics[scale=0.35]{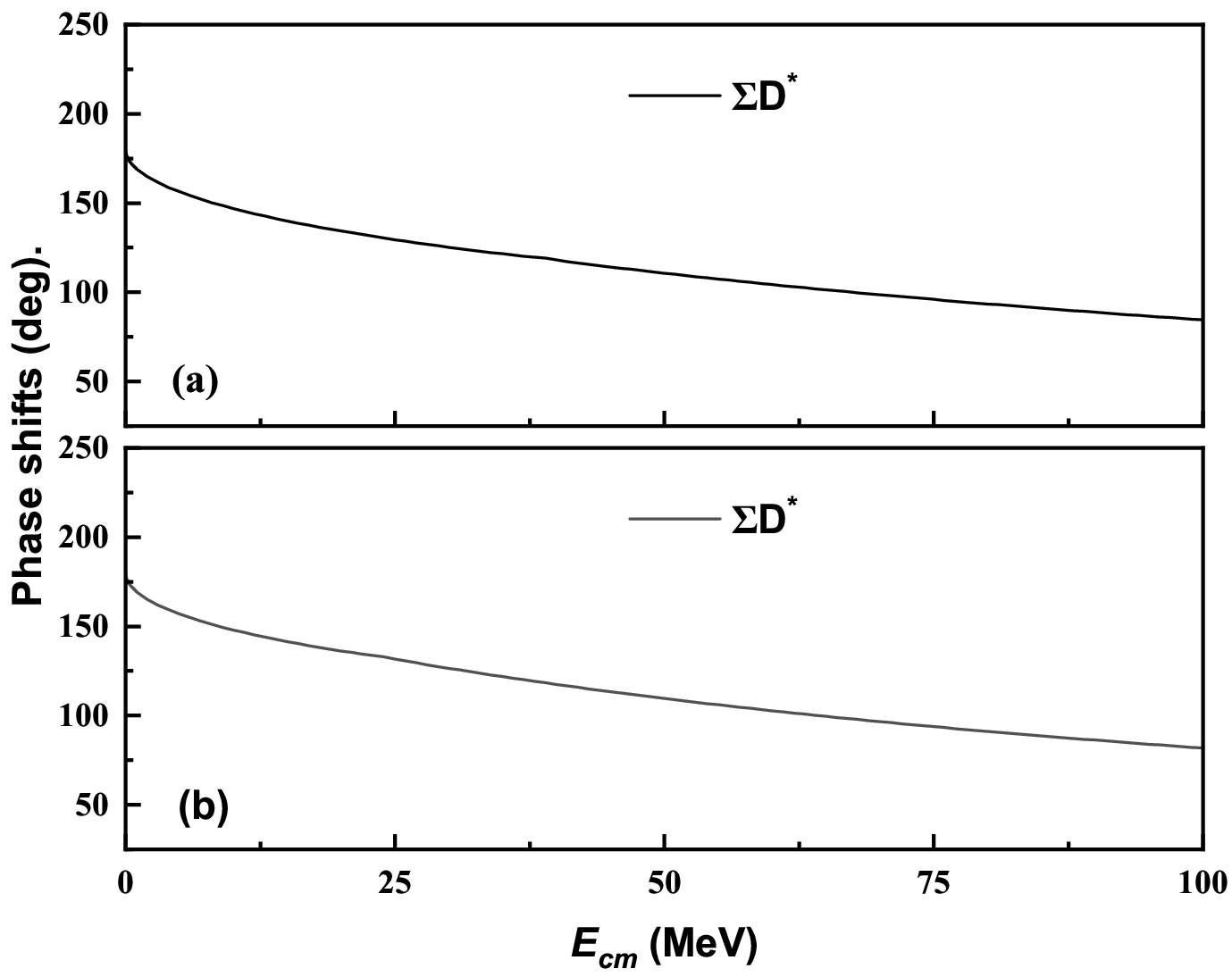}
 \caption{The phase shifts of the open channels with two channels coupling with a close channel($\Delta D_{s}$, $\Delta D_{s}^{\ast}$) and one open channel($\Sigma D^{\ast}$) for $I(J^{P})=3/2(3/2^{-})$ in QDCSM. (a) represents two channels coupling with $\Delta D_{s}^{\ast}$  and   $\Sigma D^{\ast}$, (b) denotes two two channels coupling with $\Delta D_{s}$  and   $\Sigma D^{\ast}$. }
\label{R-1.5-1.5-twocoupling}
\end{figure}

\begin{figure}
\includegraphics[scale=0.35]{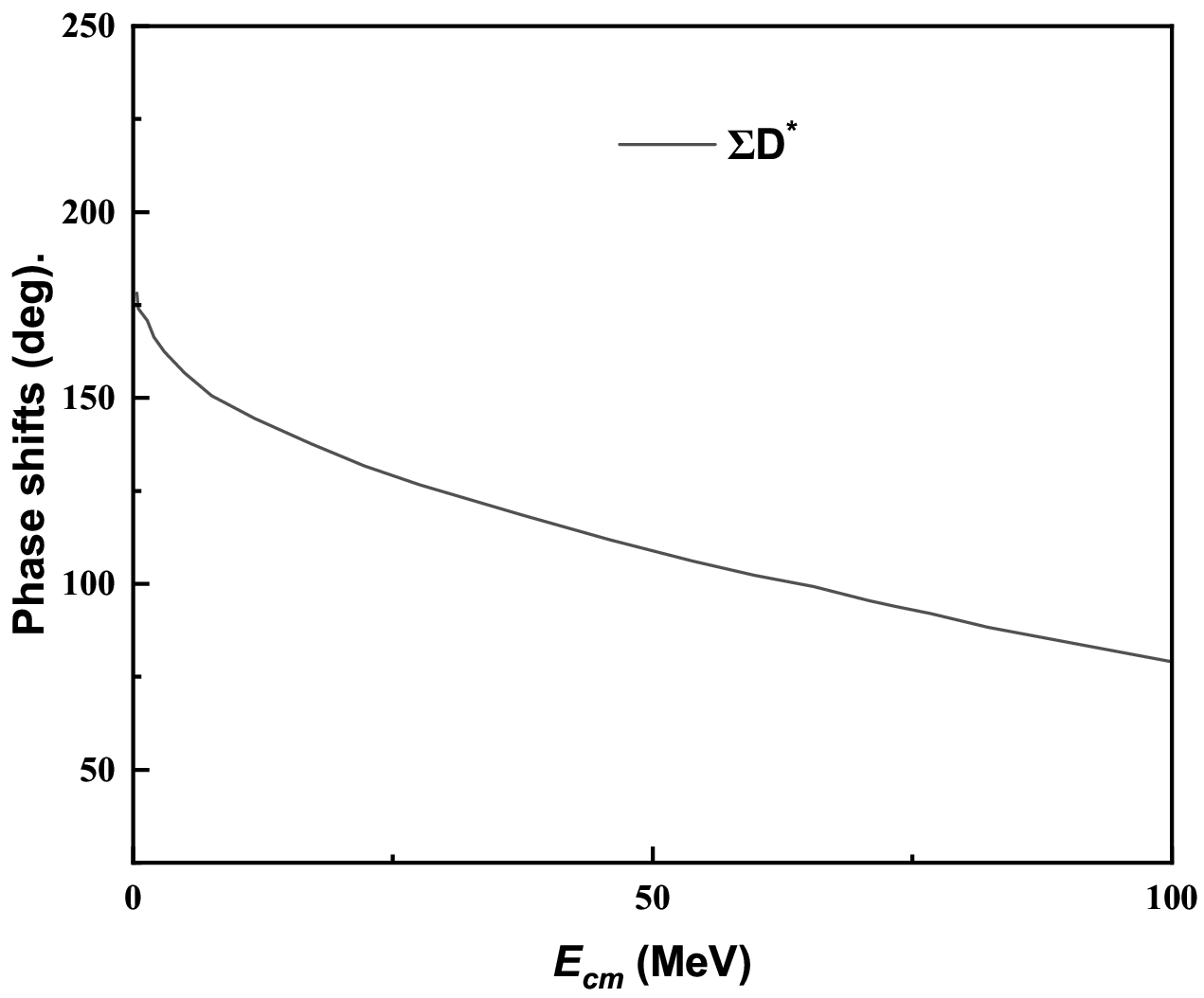}
 \caption{The phase shifts of the open channels with three channels coupling with two close channels($\Delta D_{s}$, $\Delta D_{s}^{\ast}$) and one open channel($\Sigma D^{\ast}$) for $I(J^{P})=3/2(3/2^{-})$ in QDCSM. }
\label{R-1.5-1.5-threecoupling}
\end{figure}

In the present work, several bound states are identified through single channel estimations, and their behavior when coupled to open channels with a corresponding threshold below their eigenenergies is explored. It is found that these bound states might decay into resonance states upon coupling with open channels. Specifically, if a surge in the scattering phase shifts of open channels coupling with these bound states is observed, it indicates that the bound states have transitioned into resonance states. Conversely, if no such signal is observed, the bound states may have become scattering states. To verify the existence of resonance states in the $\bar{c}snnn$ pentaquark system, all possible open channel scattering phase shifts processes are analyzed. Table~\ref{R} presents the corrected masses and decay widths of resonance states, while \cref{R-0.5-0.5-twocoupling,R-0.5-0.5-threecoupling,R-0.5-1.5-twocoupling,R-1.5-0.5-twocoupling,R-1.5-1.5-twocoupling,R-1.5-1.5-threecoupling} show the scattering processes for all possible open channels. Note that the current estimations apply only to the decay of $S$-wave open channels due to the negligible widths of the higher partial waves.


For the case of the $I(J^{P})=\frac{1}{2}(\frac{1}{2}^{-})$, two bound states, $\Sigma D$ and $\Sigma^{\ast} D^{\ast}$, emerge from the single channel dynamical estimations. Here, two types of channel coupling are explored to understand their influence on obtaining resonance states.
As shown in Fig.~\ref{R-0.5-0.5-twocoupling} and Fig.~\ref{R-0.5-0.5-threecoupling}, the first approach involves two channel coupling between one bound state and the corresponding open channel; the second approach couples all bound states with one corresponding open channel. Initially, we investigate the scenario of two channel coupling with one bound state and one different open channel. The decay channels of bound state $\Sigma D$ are $N D_{s}$ and $ND_{s}^{\ast}$, respectively.  The bound state $\Sigma^{\ast} D^{\ast}$ can decay into $N D_{s}$, $ND_{s}^{\ast}$, $\Sigma D^{\ast}$, $\Lambda D^{\ast}$ and $\Lambda D$,  respectively. Fig.~\ref{R-0.5-0.5-twocoupling} illustrates the scattering phase shifts for different open channels with two channel coupling. For the bound state $\Sigma D$, a 180-degree phase shifts surge is detected in scattering phase shifts of $N D_{s}$ and $ND_{s}^{\ast}$, indicating the formation of a resonance state $\Sigma D$. In the case of bound state $\Sigma^{\ast} D^{\ast}$,  similar resonance behavior is seen only in the scattering phase shifts process of $\Lambda D$ and $\Sigma D^{\ast}$ while no indication of the existence of resonance state  $\Sigma^{\ast} D^{\ast}$ is found in the scattering process of other open channels.  Additionally, to further explore the channel coupling effects on bound states, we have systematically studied the three channel coupling involving two bound states and one open channel. From Fig.~\ref{R-0.5-0.5-threecoupling}, only one resonance state $\Sigma D$ appears in the scattering phase shifts of $N D_{s}$ and $N D_{s}^{\ast}$, which indicates that the other bound state does not transition into a resonance state due to channel coupling but instead becomes a scattering state. This phenomenon is due to the channel coupling effect, which raises the energy of the higher state $\Sigma^{\ast} D^{\ast}$ above its threshold, causing the higher state $\Sigma^{\ast} D^{\ast}$ to vanish in the scattering processes of $N D_{s}$ and $N D_{s}^{\ast}$, respectively.

The next step is to determine the resonance masses and decay widths of obtained resonance states. The resonance mass $M_{R}^{th}$ actually is the sum of the corresponding theoretical threshold of the open channel and the incident energy at which the phase shift by $\pi/2$. The decay width of resonance states can be derived from the mass difference corresponding to the scattering phase shift at $3\pi/4$ and $\pi/4$. From Fig.~\ref{R-0.5-0.5-twocoupling} and Fig.~\ref{R-0.5-0.5-threecoupling}, the resonance mass and decay width can be inferred
from the shape of the resonance. To reduce the theoretical error and ensure the consistency of the predictions with future experimental data, we correct the resonance mass using $M_{R}=M_{R}^{th}-\sum\limits_{i} p_{i}(E_{th}^{Model}(i)-E_{th}^{Exp}(i))$, where $p_{i}$ denotes the proportion of the $i-$th channel under channel coupling effect and $M_{R}$ is the modified resonance mass. Table~\ref{R} summarises the theoretical masses, corrected masses, and decay widths corresponding to different open channels for all the attained resonance states. From Table~\ref{R},  it can be noticed that the corrected mass of resonance state $\Sigma D$ ranges from $3053$ to 3055 MeV, with a total decay width of about 13.0 to 13.4 MeV. The resonance state $\Sigma^{\ast} D^{\ast}$ has a mass range of 3389$\sim$3390 MeV and a total decay width of 10.4 MeV.

In addition, the low energy scattering behavior shown in Fig.~\ref{R-0.5-0.5-twocoupling} and Fig.~\ref{R-0.5-0.5-threecoupling} provides important information. For instance, the low-energy scattering phase shifts of $ND_{s}$ and $ND_{s}^{\ast}$ converge to 180 degrees as the incident energy approaches zero but decay rapidly with increasing incident energy, reflecting the formation of bound states in $ND_{s}$ and $ND_{s}^{\ast}$ due to channel coupling. In contrast, the low energy scattering phase shifts of $\Sigma D^{\ast}$, $\Lambda D^{\ast}$, and $\Lambda D$ indicate that these channels cannot form bound states through channel coupling.

For the case of the $I(J^{P})=\frac{1}{2}(\frac{3}{2}^{-})$, only a bound state $\Sigma^{\ast} D$ in the single channel estimations is identified. This bound state can decay into open channels $ND^{\ast}_{s}$, $\Sigma D^{\ast}$, and $\Lambda D^{\ast}$, respectively. In the current research, two channel coupling between an open channel and the bound state is employed to explore the presence of resonance state $\Sigma^{\ast} D$ in the scattering phase shifts of certain open channels. As shown in Fig.~\ref{R-0.5-1.5-twocoupling}, the scattering phase shifts for all possibility open channels are presented, revealing resonances state $\Sigma^{\ast} D$ in the scattering phase shifts of open channels $\Sigma D^{\ast}$ and $\Lambda D^{\ast}$, but not in open channel $ND^{\ast}_{s}$. Additionally, the corrected resonance mass and decay width of the resonance state $\Sigma^{\ast} D$ are reported in Table~\ref{R} as 3250$\sim$3252 MeV and 4.4 MeV, respectively. For the state with $I(J^{P})=\frac{1}{2}(\frac{5}{2}^{-})$, there is only one channel $\Sigma^{\ast} D^{\ast}$. Based on the bound state estimation results, the state $\Sigma^{\ast} D^{\ast}$ is a scattering state if only the $S$-wave is considered.

In the $\bar{c}snnn$ pentaquark states with $I=\frac{3}{2}$, Figs.~\ref{R-1.5-0.5-twocoupling},~\ref{R-1.5-1.5-twocoupling} ,and~\ref{R-1.5-1.5-threecoupling} illustrate the scattering phase shifts behavior for all possible open channels. The results indicate that regardless of whether two channels or three channels coupling is considered, no evidence of resonance states is observed in the corresponding scattering phase shifts, which suggests that the bound states obtained from single channel estimations cannot transition into resonance states via channel coupling but rather become scattering states.  In the $\bar{c}snnn$ pentaquark state with $J^{P}=\frac{1}{2}^{-}$, the bound state $\Delta D^{\ast}$ can decay into $\Sigma D$ and $\Sigma D^{\ast}$, respectively. Analysis of the scattering phase shifts for $\Sigma D$ and $\Sigma D^{\ast}$ shows no resonance state $\Delta D^{\ast}$, indicating that the bound state $\Delta D^{\ast}$ becomes a scattering state. For the $\bar{c}sqqq$ pentaquark state with $J^{P}=\frac{3}{2}^{-}$, the results are unfortunately similar to those with $J^{P}=\frac{1}{2}^{-}$. No resonance states are detected in the scattering phase shifts of open channel $\Sigma D^{\ast}$ under two channels coupling and three channels coupling, as shown in Fig.~\ref{R-1.5-1.5-twocoupling} and~\ref{R-1.5-1.5-threecoupling}. Finally, for the $\bar{c}snnn$ pentaquark state with $J^{P}=\frac{5}{2}^{-}$, although a bound state $\Delta D^{\ast}$ is identified, it cannot decay into $\Sigma^{\ast} D^{\ast}$ in the $S$-wave condition because its threshold is below that of $\Sigma^{\ast} D^{\ast}$.

\subsection{The magnetic moments of the $\bar{c}snnn$ pentaquark states}

\begin{table}[t]
    \begin{center}
        \renewcommand{\arraystretch}{1.5}
        \caption{\label{mag-mom} The magnetic moments of the $\bar{c}snnn$ pentaquark states with different quantuma numbers. The magnetic moments are given in units of $\mu_N$. }
        \begin{tabular}{p{1.5cm}<\centering p{3.5cm}<\centering p{1.5cm}<\centering p{1.5cm}<\centering p{1.5cm}<\centering   }
            \toprule[1pt]
            $I(J^{P})$  & Channel &  $\mu_{Channel}$\\
            \midrule[1pt]
            $\frac{1}{2}(\frac{1}{2}^{-})$ &$\Sigma D$                   &2.17976   \\
                                           &$\Sigma^{\ast} D^{\ast}$     &1.2373   \\
                                           & all channel coupling        &1.99787   \\
            $\frac{1}{2}(\frac{3}{2}^{-})$ & $\Sigma^{\ast} D$           &2.45114   \\
                                           & all channel coupling        &1.07887   \\
            $\frac{3}{2}(\frac{1}{2}^{-})$ & $\Delta D_{s}^{\ast}$       &3.63612   \\
            $\frac{3}{2}(\frac{3}{2}^{-})$ & $\Delta D_{s}$              &5.99361   \\
                                           & $\Delta D_{s}^{\ast}$       &4.02771   \\
                                           & all channel coupling        &4.47025   \\
            \bottomrule[1pt]
        \end{tabular}
    \end{center}
\end{table}

The aforementioned studies mainly focus on the mass spectrum and decay properties of the anticharmed-strange pentaquark states. Although these data provide some information, they are still insufficient for a thorough comprehension of the internal structure of the anticharmed-strange pentaquark states. To better uncover the internal composition of these exotic particles, we will further investigate the magnetic moments of the anticharmed-strange pentaquark states. Research on magnetic moments can reveal the charge distribution and reflect the geometric configuration of hadrons, thus offering a new perspective for understanding the internal structure of these complex particles. Given that this study is primarily carried out in the $S$-wave system, the contribution of the orbital angular momentum can be neglected in the determination of the magnetic moments. The magnetic moment of the anticharmed-strange pentaquark states can be written as
\begin{equation*}
    \mu=\langle \psi_{IJ}| \hat{\mu}_m | \psi_{IJ} \rangle,
\end{equation*}
where $\hat{\mu}_m=\sum^{5}_{i=1}\frac{\hat{Q}_{i}}{2m_{i}}\sigma_{i}^{\mathcal{Z}}$ is the magnetic moment operator, $\hat{Q}_{i}$ is the electric charge operator of the $i-$th quark, $\sigma_{i}^{\mathcal{Z}}$ is the third component ($\mathcal{Z}-component$) of Pauli matrix, and $\psi_{IJ}$ is the eigenvector of those states.

Table~\ref{mag-mom} details the magnetic moments of the anticharmed-strange pentaquark states across different quantum numbers, expressed in units of $\mu_{N}$. For the $I(J^{P})=\frac{1}{2}(\frac{1}{2}^{-})$ state, the magnetic moments are 2.17976 for the $\Sigma D$ channel, 1.2373 for the $\Sigma^{\ast}D^{\ast}$ channel, and 1.99787 for all channel coupling. For the $I(J^{P})=\frac{1}{2}(\frac{3}{2}^{-})$ state, the obtained values are 2.45114 for the $\Sigma^{\ast} D$ channel and 1.07887 for all channel coupling. For the $I(J^{P})=\frac{3}{2}(\frac{1}{2}^{-})$ and $\frac{3}{2}(\frac{3}{2}^{-})$  state,  the gained values for different channels and all channel coupling are provided, with the highest magnetic moment being 5.99361 for the $\Delta D_{s}$ channel in the $\frac{3}{2}(\frac{3}{2}^{-})$ state and the lowest being  3.63612 for the $\Delta D_{s}^{\ast}$ channel in the $\frac{3}{2}(\frac{1}{2}^{-})$ state. These data will provide detailed insights into the magnetic moments' properties of various anticharmed-strange pentaquark states.

\section{Summary\label{sum}}
In the present study, we systematically investigate anticharmed-strange pentaquark states with quark composition $\bar{c}snnn$ using the Resonating Group Method within the QDCSM framework. To explore the existence of these states, we estimate effective potential to confirm interactions for different quantum numbers. Additionally, both single channel and multi-channel coupling estimations are performed to hunt for possible bound states. Our estimations suggest the existence of three bound states for the anticharmed-strange pentaquark system with $I(J^{P})$ quantum numbers of $\frac{1}{2}(\frac{1}{2}^{-})$, $\frac{1}{2}(\frac{1}{2}^{-})$ and $\frac{1}{2}(\frac{3}{2}^{-})$, with estimated masses of 2886 MeV, 3039 MeV and 3153 MeV, respectively.

Alternatively, various open channel scattering phase shifts are analyzed to search for possible resonance states. The consequences of the investigations demonstrate the occurrence of three resonance states within the anticharmed-strange pentaquark system. Two resonance states are detected with the quantum number $\frac{1}{2}(\frac{1}{2}^{-})$. The resonance state $\Sigma D$, with mass 3053$\sim$3055 MeV and decay width 13.0$\sim$13.4 MeV, is identified in the scattering phase shifts of $ND_{s}$ and $N D_{s}^{\ast}$. The other resonance state, $\Sigma^{\ast} D^{\ast}$, is spotted in the $\Sigma D^{\ast}$ and $\Lambda D$ channel, with a mass of 3389$\sim$3390 MeV and decay width of 10.4 MeV. Apart from the two resonance states that have already been found as discussed above, another resonance state, $\Sigma^{\ast} D$, with a quantum number of $\frac{1}{2}(\frac{3}{2}^{-})$, is visible in the scattering shifts involving $\Lambda D^{\ast}$ and $\Sigma D^{\ast}$, with a mass of 3250$\sim$3252 MeV and decay width of 4.4 MeV. Additionally, we also provide information on the magnetic moments to better understand the internal structure of the obtained explicit pentaquark states.

\acknowledgments{This work is supported partly by the National Natural Science Foundation of China under Contract Nos. 12175037, 12335001, 11775118, 11535005, 12205249, 12205125, and 11865019, and is also supported by the Fundamental Research Funds for the Central Universities No. 2242022R20040; and the China Postdoctoral Science Foundation No. 2021M690626 and No. 1107020201. Besides, Jiangsu Provincial Natural Science Foundation Project (No. BK20221166), National Youth Fund: No. 12205125 and School-Level Research Projects of Yancheng Institute of Technology (No. xjr2022039) also supported this work.}

\bibliography{uusucbar}

\end{document}